\documentclass{PoS}
\usepackage{amssymb}    
\usepackage{amsmath}
\usepackage{wrapfig}

\title{$B_c \to J/\psi(B_s,B_s^*)+ n\pi$ Decays}

\ShortTitle{$B_c \to J/\psi(B_s,B_s^*)+ n\pi$ Decays}

\author{\speaker{Alexander Berezhnoy}\thanks{This research is partially supported by Russian Foundation for Basic Research (grant 10-02-00061a).}\\
        SINP MSU, Moscow, Russia\\
        E-mail: \email{Alexander.Berezhnoy@cern.ch}}

\author{Anatolii Likhoded\\
        IHEP, Provtino, Russia\\
        E-mail: \email{Anatolii.Likhoded@ihep.ru}}

\author{Alexey Luchinsky\thanks{The work of A.~V.~Luchinsky was partially supported by non-commercial foundation ''Dynasty'' and the grant of the president of Russian Federation for young scientists with PhD degree (grant MK-406.2010.2).}\\
        IHEP, Provtino, Russia\\
        E-mail: \email{alexey.luchinsky@gmail.com}}

\abstract{The $B_c \to J/\psi(B_s,B_s^*)+ n\pi$ amplitudes are calculated 
in the frame work of factorization model, which allows to represent this processes as  $B_c$ decay into  $J/\psi (B_s)+W^*$ followed by the virtual $W^*$-boson decay into the final set of $\pi$-mesons. The module for the generation of $B_c$ meson decays into $J/\psi  + n\pi$ and $B_s^{(*)}  + n\pi$  ($n\le 3$)  is implemented into the GAUSS program package of the LHCb Collaboration. }

\FullConference{The XXth International Workshop High Energy Physics and Quantum Field Theory \\
		September 24-October 1, 2011\\
		Sochi Russia}

\begin{document}

\section{Introduction}

The measurements of $B_{c}$ meson mass and lifetime in CDF \cite{Aaltonen:2007gv}
and D0  \cite{Abazov:2008rb} experiments  were the first steps in the experimental research of heavy quarkonia with open flavor. The obtained experimental results  are in a good agreement with the theoretical predictions for the $B_c$ mass~\cite{Gershtein1995aaa, Eichten:1994gt, Ebert:2002pp}:
$$m_{B_c}^{\rm CDF}=6.2756\pm 0.0029(\textrm{stat.})
\pm 0.0025(\textrm{sys.}) \;\textrm{GeV}, $$
$$m_{B_c}^{\rm D0}=6.3000\pm 0.0014(\textrm{stat.})
\pm 0.0005(\textrm{sys.}) \;\textrm{GeV},
$$
$$ \qquad  m_{B_c}^{\rm theor}=6.25\pm 0.03 \;\textrm{GeV};$$
as well as for the decay time~\cite{Kiselev:1999sc,Kiselev:2000pp,Kiselev:1994ay,Kiselev:1993eb,Kiselev:1992tx}:
$$\tau_{B_c}^{\rm CDF}=0.448^{+0.038}_{-0.036} (\rm stat.)\pm 0.032 (\rm sys.)\;\textrm{ps},$$
$$\tau_{B_c}^{\rm D0}=0.475^{+0.053}_{-0.049} (\rm stat.)\pm 0.018 (\rm sys.)\;\textrm{ps},$$
$$\tau_{B_c}^{\rm theor}=0.48 \pm 0.05 \;\textrm{ps}.$$

The $B_c$ meson is observed at Tevatron only in two decay modes: $B_c\to J/\psi \pi$ and
$B_c\to J/\psi + \mu +\nu_{\mu}$. The investigation of other decay modes is possible at
at LHC, where about $10^{10}$ events with $B_c$ mesons per year are expected. 
This huge amount of events will allow to obtain the information on the production cross section distributions, on the decay branching fractions, and in some cases, on the distributions of decay products. And absolutely new results on $B_c$ decays have already obtained at LHC: the $B_c$ decay to $J/\psi +3\pi$ has been first observed by LHCb collaboration~\cite{Skwarnicki}.

The  $B_c$ systems do not have strong and electromagnetic annihilation decay modes. Due to this reason the exited $B_c$ systems laying below $B+D$ threshold have the decay widths by two order of magnitude smaller  than  the widths for analogous exited states of charmonium and bottonium. All exited $B_c$ states after a set of radiative transitions decay into the lightest pseudoscalar state ($0^-$). The lifetime of this state is comparable with  the lifetimes of  
of $B$ and $D$ mesons and essentially differ from lifetimes of other lightest quarkonia: $\eta_c$   and $\eta_b$.  This is why  $B_c$ meson 
provides a unique possibility to investigate the both  strong and weak interactions.

The main  $B_c$ decay modes, such as  $B_c\to J/\psi+\ell\nu$, $B_c\to J/\psi\pi$ and $B_c\to B_s^{(*)}\rho$  were theoretically studied in details (see, for example \cite{Kiselev:1999sc,Kiselev:2002vz,Wang:2008xt}). For all these processes it is assumed that the factorization approach is valid: the  decay $B_c \to \textrm{heavy  hadron} +W^*$  is followed by the decay of  the virtual $W$-boson. The transition form-factors for the processes $B_c\to \textrm{heavy hadron} +W^*$ can be estimated within QCD sum rules, within quark potential models,  or in the framework of light-cone quark model.

In our recent articles \cite{Likhoded:2009ib, Likhoded:2010jr} we have studied the $B_c$ decay process with several pions in the final state, such as   $B_c\to
J/\psi+n\pi$ and $B_c\to B_s^{(*)}+n\pi$ with $1\le n\le 4$. The factorization approach have been used in these calculations. The characteristics of the virtual $W$-boson decay have been adopted from the experimental data on $\tau\to\nu_\tau+n\pi$ decays for $n\le 3$  and from the experimental data on $\pi$ mesons production in the process $e^+e^-\to 4\pi$.

The modules for calculation of these decay amplitude are implemented into EvtGen program package~\cite{EvtGen}.
Therefore the detailed simulation of these decays in the kinematical condition of real experiments is possible now. In this paper we describe the installation and using the developed modules.

\section{Calculation technique}

\begin{wrapfigure}{r}{0.5\textwidth}
\vspace*{-0.5cm}
\begin{center}
\includegraphics[width=0.5\textwidth]{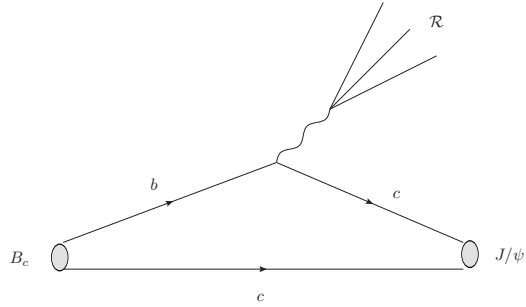}
\end{center}
  \caption{Typical diagram for $B_c\to V(P)+n\pi$ decay.}
  \label{fig:diag}
\end{wrapfigure}
To calculate the decay amplitudes we  assume that the discussed processes can be 
represented as  the  decay $B_c\to \textrm{heavy  hadron} +W^*$  followed by the decay of  the virtual $W$-boson. An additional assumption should be made  that the only one of the constituent quarks decays weakly, meanwhile the other quark remains the same (a typical diagram is presented in Fig.~\ref{fig:diag}).

Within this approach the amplitude of the process  can be written in the form
\begin{equation}
  \mathcal{A}[B_c \to  J/\psi + n\pi] = \frac{G_F V_{cb}}{\sqrt{2}} a_1(\mu_R) \mathcal{H}^{J/\psi}_\mu \epsilon_W^\mu,
  \label{eq:amp_Jpsi}
\end{equation}
where $\epsilon_W$ is the polarization vector of $W^*$;

$\mathcal{H}^{J/\psi}$ is the $B_c\to J/\psi +
W^*$ transition vertex:
\begin{equation}
\mathcal{H}_{\mu} = \left\langle J/\psi\left|\bar{c}\gamma_{\mu}\left(1-\gamma_{5}\right)b\right|B_{c}\right\rangle =\mathcal{V}_{\mu}-\mathcal{A}_{\mu}.
\end{equation}

Vector and axial currents are equal to
\begin{equation}
\mathcal{V}_{\mu}  =  \left\langle J/\psi\left|\bar{c}\gamma_{\mu}b\right|B_{c}\right\rangle =
i\epsilon^{\mu\nu\alpha\beta}\epsilon_{\nu}^{\psi} p_{\alpha}q_{\beta}F_{V}\left(q^{2}\right),
\end{equation}

\begin{multline}
\mathcal{A}_{\mu}  = \left\langle J/\psi\left|\bar{c}\gamma_{\mu}\gamma_{5}b\right|B_{c}\right\rangle = \\
\epsilon_{\mu}^{\psi}F_{0}^{A}\left(q^{2}\right)+p_{\mu}\left(\epsilon^{\psi}p_{B_c}\right)F_{+}^{A}\left(q^{2}\right)+q_{\mu}\left(\epsilon^{\psi}p_{B_c}\right)F_{-}^{A}\left(q^{2}\right),
\end{multline}
where $p_{B_c}$ and $p_{J/\psi}$ are the momenta of $B_{c}$- and $J/\psi$-mesons;

$q= p_{B_c}-p_{J/\psi}$ is the momentum of virtual $W$-boson;

 $p= p_{B_c}+p_{J/\psi}$;

$\epsilon_\mu^{J/\psi}$ is the polarization vector of $J/\psi$ meson;

and $F_{V}(q^{2})$, 
 $F^A_{0}(q^2)$, $F^A_{+}(q^2)$, $F^A_{-}(q^2)$ and $F_V(q^2)$ are form-factors of $B_c\to J/\psi+W^*$ decays.

In the tree approximation the parameter $a_1(\mu_R)$  is equal to unity. Higher-order corrections lead to dependence of this factor on
renormalization scale $\mu_R$ \cite{Buchalla:1995vs}. Numerical values for $a_1(\mu_R)$ at different scale are  calculated in \cite{Kiselev:2002vz} .
For the process   $B_c\to J/\psi+n\pi$  the value of $\mu_R$ has been chosen to be equal to the  mass value of the decayed $b$-quark:
$$a_1(m_b) = 1.4$$

The form-factors were calculated within different nonperturbative approaches: QCD sum rules \cite{Kiselev:2002vz}, potential quark models \cite{Kiselev:1993eb}, and
Light-Front quark models \cite{Anisimov:1998uk, Huang:2008zg}. In our article we use exponential parametrization of these form-factors (see Tab.~\ref{tab:ff}):
\begin{eqnarray}
  F_i(q^2) &=& F_i(0) \exp\left\{ -c_1 q^2 - c_2 q^4\right\}.
  \label{eq:Fi}
\end{eqnarray}

\begin{wraptable}{r}{0.35\textwidth}
\begin{center}
\vspace*{-0.5cm}
\begin{tabular}{|c|c|c|c|}
\hline 
\multicolumn{4}{|c|}{$B_c \to J/\psi +n \pi$} \\ \hline
 &$F_i(0)$ & $c_1$ & $c_2$  \\ \hline
 $A_0$ & 5.9 & 0.049 & 0.0015 \\ \hline
 $A_+$ &-0.074 & 0.049 & 0.0015  \\ \hline
 $V$ &	0.11 & 0.049 & 0.0015 \\ \hline  \hline
\multicolumn{4}{|c|}{$B_c \to B_s^* +n \pi$} \\ \hline
 &$F_i(0)$ & $c_1$ & $c_2$  \\ \hline
 $A_0$ & 8.1 & 0.30 & 0.069 \\ \hline
 $A_+$ &0.15 & 0.30 & 0.069  \\ \hline
 $V$ &	1.08 & 0.30 & 0.069 \\ \hline \hline
\multicolumn{4}{|c|}{$B_c \to B_s +n \pi$} \\ \hline
 &$f_i(0)$ & $c_1$ & $c_2$  \\ \hline
 $f_+$ & 1.3  & 0.30  & 0.069   \\ \hline
\end{tabular}
\end{center}
\caption{Form-factor parameters for different SR form-factor sets.}
\label{tab:ff}
\end{wraptable}

The decay  $B_c\to B_s^*+n\pi$ is described by analogy with the decay $B_c\to J/\psi+n\pi$. The same formula for the amplitude is used with the other parameter set (see Tab.~\ref{tab:ff}).

The amplitude for the decay  $B_c\to B_s+n\pi$   can be written in more simple form
\begin{eqnarray}
  \mathcal{A}[B_c \to  B_s + n\pi] &=& \frac{G_F V_{cs}}{\sqrt{2}} a_1(\mu_R) \mathcal{H}^{B_s}_\mu \epsilon_W^\mu,
  \label{eq:amp_Bs}
\end{eqnarray}
where
\begin{equation}
  \mathcal{H}^{B_s}_\mu = \left\langle B_s\left|\bar{c}\left(1-\gamma_{5}\right)b\right|B_{c}\right\rangle
 = f_+(q^2) p_\mu + f_-(q^2) q_\mu.
\end{equation}

The detailed information about $W^*$ decay is not needed  to obtain the integrated branching fractions, as well as the branching fraction distributions on $q^2$. Let us consider the decay 
$B_c\to J/\psi W^* \to J/\psi n \pi$ as an example:

\begin{equation}
d\Gamma\left(B_c\to J/\psi\mathcal{R}\right)  = \\
 \frac{1}{2M}\frac{G_F^{2}V_{cb}^2}{2}a_1^2\mathcal{H}^{\mu}\mathcal{H}^{*\nu}
{\epsilon_{\mu}^W}^*{\epsilon_{\nu}^W}d\Phi\left(B_c\to J/\psi n \pi\right),
\end{equation}

where Lorentz-invariant phase space is defined according to

\begin{equation}
d\Phi\left(Q\to p_{1}\dots p_{n}\right) =  (2\pi)^{4}\delta^{4}\left(Q-\sum p_{i}\right)\prod\frac{d^{3}p_{i}}{2E_{i}(2\pi)^{3}}.
\end{equation}

Using the following recurrent expression for the phase space
\begin{equation}
d\Phi\left(B_{c}\to J/\psi W^* \to J/\psi n \pi \right)  =  \frac{dq^{2}}{2\pi}d\Phi\left(B_{c}\to J/\psi W^{*}\right)d\Phi\left(W^{*}\to n \pi \right)
\end{equation}
 one can perform the integration over phase space of the final state $n \pi$:
\begin{equation}
\frac{1}{2\pi}\int d\Phi\left(W^{*}\to n \pi \right)\epsilon_{\mu}^{W}{\epsilon_{\nu}^W}^*  =  \left(q_{\mu}q_{\nu}-q^{2}g_{\mu\nu}\right)\rho_{T}\left(q^{2}\right)+q_{\mu}q_{\nu}\rho_{L}\left(q^{2}\right),
\end{equation}
where spectral functions $\rho_{T,L}\left(q^{2}\right)$
are universal and can be determined from theoretical and experimental
analysis of some other processes, for example $\tau\to\nu_{\tau} n \pi$
decay or electron-positron annihilation $e^{+}e^{-}\to n \pi$. (See
\cite{Likhoded:2009ib,Likhoded:2010jr,Schael:2005am} for details.)
It is worth to mention, that due to the vector current conservation and the partial axial current conservation
spectral function $\rho_{L}$ for $n\geq 2$ is negligible in almost
whole kinematical region, so  it can be neglected in the estimations for $n\geq 2$.
 For the purposes of our article, however, a more detailed description is
required for the multipion final state.

\begin{figure}[!t]
\centering
\resizebox*{1.0\textwidth}{!}{  
\includegraphics{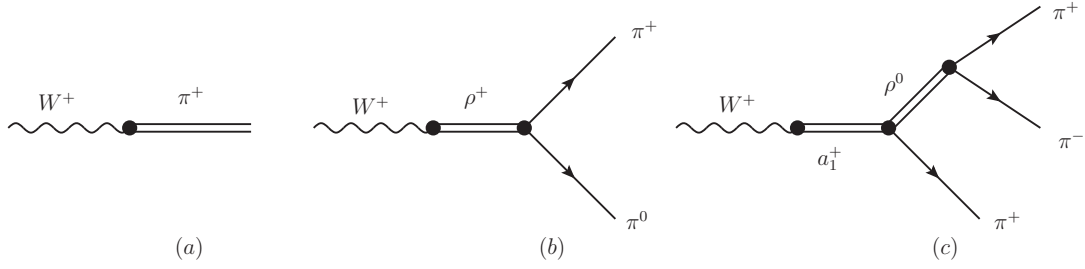}
}
  \caption{Typical diagrams for $W^*\to n\pi$ decay in resonance approximation.}
  \label{fig:digPi}
\end{figure}

In the framework of resonance model the decays of $W^*$-boson can be described in terms of virtual $\rho$- and $a_1$-mesons exchange
(see typical diagrams presented in Fig.~\ref{fig:digPi}).

If there is one $\pi$-meson in the final state (see Fig.~\ref{fig:digPi}a), the vertex of $W^*\to\pi$ transition can be written in the form
\begin{equation}
  \langle \pi^+ |(\bar d u)_{V-A} | W\rangle = f_\pi k_\mu,
\end{equation}
where $k$ is the momentum of $\pi$-meson and  $f_\pi\approx 140$ MeV is its coupling constant. In accordance with this interaction vertex the effective polarization vector $\epsilon^W_\mu$ in this case has the form
\begin{equation}
  \epsilon^W_\mu = f_\pi k_\mu/m_\pi^2.
\end{equation}
Note, that this vertex violates the axial current.

 The decays of virtual $W$-boson into multipion final states  are described within resonance model in terms of virtual $\rho$- and $a_1$-mesons exchange (see typical diagrams presented in Fig.~\ref{fig:digPi}b,c).

The $W^*\to\pi^+\pi^0$ decay is saturated mainly by contributions of virtual $\rho$-
and $\rho'$-mesons (see Fig.~\ref{fig:digPi}b). The corresponding effective polarization vector can be written as
\begin{equation}
  \epsilon^{2\pi}_\mu = F_\rho(q^2) (k_1-k_2)_\mu,
  \label{eq:eps2pi}
\end{equation}
where $k_{1,2}$ are $\pi$-mesons momenta and $F_\rho(q^2)$ is the $\rho$-meson form-factor (see  \cite{Kuhn:1990ad}). The difference in $\pi^0$- and $\pi^+$-meson masses is neglected, thus the virtual $W$ boson in this decay has a transverse polarization. It should be noted that the width of the $\rho$ meson must be taken into account.

The $W^*\to 3\pi$-transition is saturated mainly by $W^*\to a_1\to\rho\pi\to3\pi$
decay chain. Following \cite{Kuhn:1990ad} one can write the effective polarization vertex in this case as
\begin{equation}
  \epsilon^{3\pi}_\mu = -i \frac{2\sqrt{2}}{3f_\pi}F_a(q^2)\left\{
    B_\rho(s_2) V_{1\mu} + B_\rho(s_1) V_{2\mu}
  \right\},
  \label{eq:eps3pi}
\end{equation}
where
\begin{equation}
  V_{1,2\mu} = k_{1,2\mu}-k_{3\mu}-q_\mu \frac{q(k_{1,2}-k_3)}{q^2}
\label{eq:V3pi}
\end{equation}
and
\begin{equation}
  s_{1,2} = (q-k_{1,2})^2.
\end{equation}
Parametrization of $B_\rho(s)$ function is presented in \cite{Kuhn:1990ad}. It can be clearly seen, that if one neglects the difference between charged and neutral $\pi$-meson masses the above expression in transverse and the axial current is conserved.  In EvtGen package this transition was realized already in TAUHADNU model.

\newpage
\section{Software structure\label{sec:soft}}

\begin{wrapfigure}{r}{0.6\textwidth}
\centering
\includegraphics[width=0.6\textwidth]{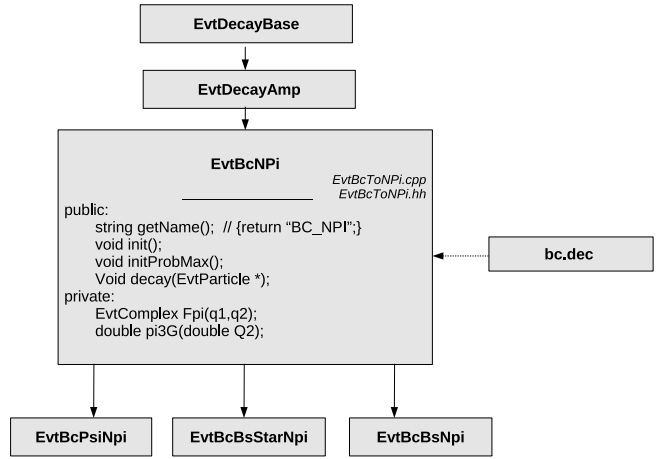}
  \caption{The program structure.}
  \label{fig:class}
\end{wrapfigure}

For the Monte-Carlo simulation of the discussed decays the generator EvtGen is used, which is the part of LHCb software environment  GAUSS. 
 EvtGen is written completely on C++. In order to add a new decay to it one should just create a new class, that describes this decay. The base class is \texttt{EvtDecayAmp}, where prototypes of all necessary for the model functions are given. The most significant of these functions are \texttt{init()}, \texttt{initProbMax()} and \texttt{decay()} (see Fig.~\ref{fig:class}).

The method \texttt{init()} performs the initialization of the decay model and reads its parameters. The necessary parameters are stored in the so-called .dec-file (the exact name of this file can be determined by the user). The detailed description of the .dec-file format can be found in the EvtGen documentation~\cite{EvtGen}. 

\begin{figure}[!b]
\centering
\resizebox*{1.\textwidth}{!}{
\includegraphics[width=1.4\textwidth]{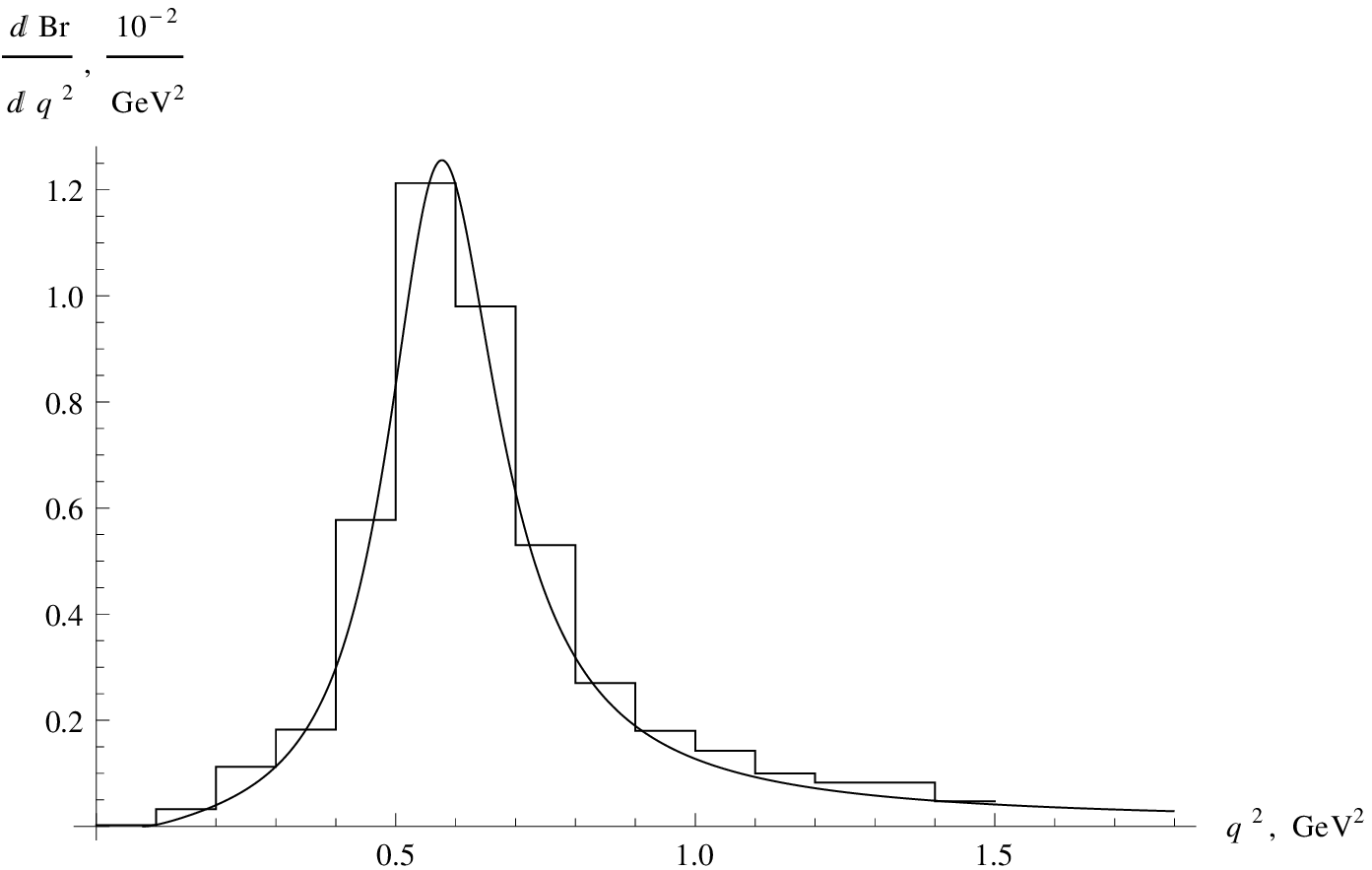}
\hfill
\includegraphics{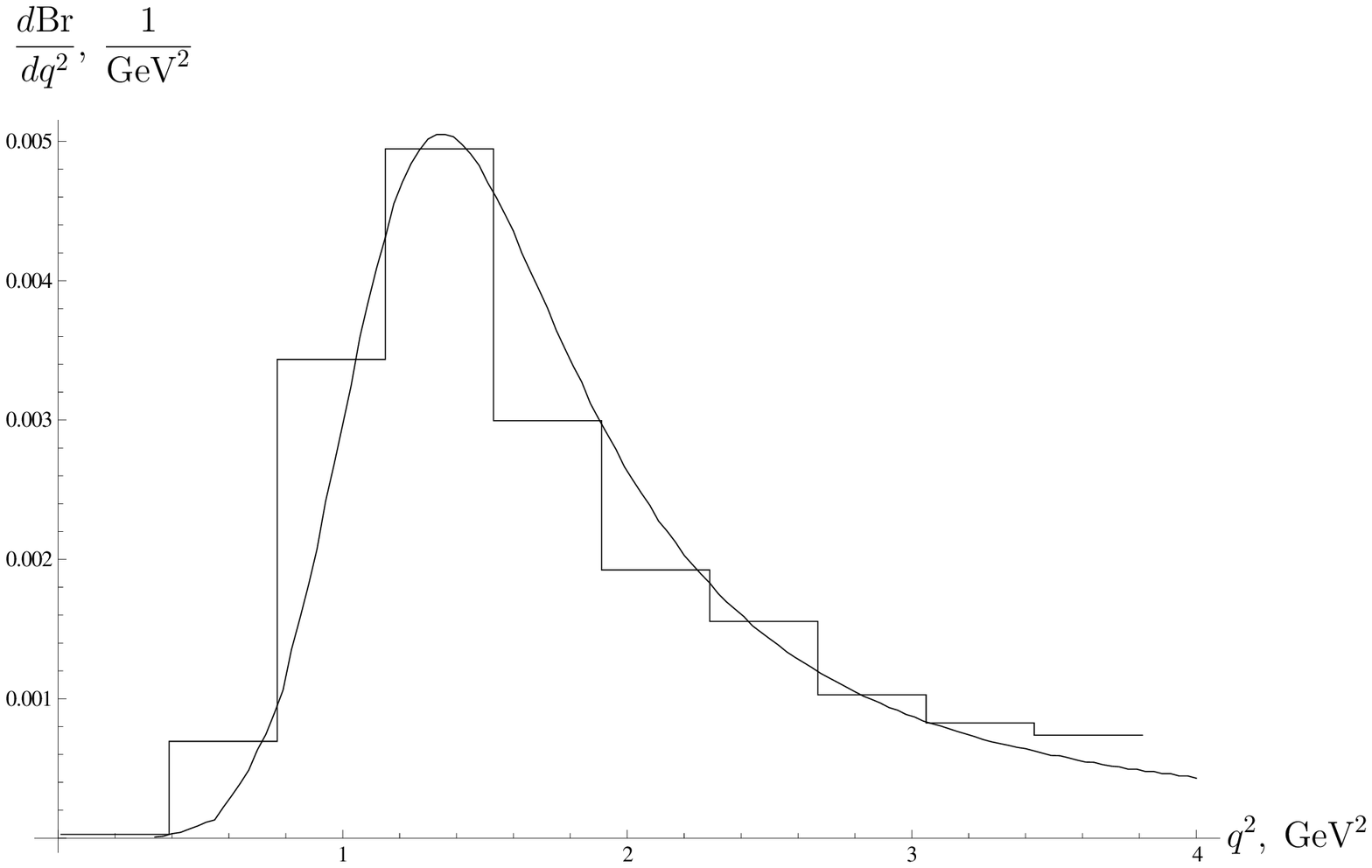}
}
\\
\parbox[t]{0.45\textwidth}{
\caption{The distribution over the squared invariant mass $q^2=m_{\pi\pi}^2$ in $B_c\to J/\psi+2\pi$ decay generated within BC\_NPI-model (histogram) in comparison with the predictions~\cite{Likhoded:2009ib}.  \hfill}
\label{fig:comp}
}
\hfill
\parbox[t]{0.45\textwidth}{
\caption{The distribution over the squared invariant mass $q^2=m_{\pi\pi\pi}^2$ in $B_c\to J/\psi+3\pi$ decay generated within BC\_NPI-model (histogram) in comparison with the predictions~\cite{Likhoded:2009ib}.\hfill}
\label{fig:comp_1}
}

\end{figure}

As it is seen from Fig.~\ref{fig:comp} and \ref{fig:comp_1} the results of Monte-Carlo simulation for the decays
$B_c\to J/\psi + 2\pi$ and $B_c\to J/\psi + 3\pi$ are in good agreement with the theoretical predictions~\cite{Likhoded:2009ib}.

\section{Results}

\begin{itemize}
  \item The branching ratios for the decays $B_c \to J/\psi(B_s,B_s^*) + n\pi$ have been estimated in the framework of the factorization approach.

\item The width of intermediate mesons must be taken into account.

  \item The package for simulation of these decays within the standard LHCb software GAUSS has been developed.

  \item The experimental LHCb data on $B_c \to J/\psi +3 \pi$ decay~\cite{Skwarnicki} can be described satisfactorily within the model.

\end{itemize}


\begin{thebibliography}{99}
\bibitem{Aaltonen:2007gv}
T.~Aaltonen \emph{et~al.}, Phys. Rev. Lett. \textbf{100}, 182002 (2008).

\bibitem{Abazov:2008rb} 
V.~M. Abazov \emph{et~al.}, Phys. Rev. Lett. \textbf{102}, 092001 (2009).

\bibitem{Gershtein1995aaa} 
S. S. Gershtein,  V. V. Kiselev, A. K. Likhoded, and A. V. Tkabladze, Phys.Rev.D \textbf{51}, 3613 (1995).

\bibitem{Eichten:1994gt}
  E.~J.~Eichten and C.~Quigg,
  Phys.\ Rev.\  D {\bf 49}, 5845 (1994).

\bibitem{Ebert:2002pp}
  D.~Ebert, R.~N.~Faustov and V.~O.~Galkin,
  Phys. Rev. D {\bf 67}, 014027 (2003).

\bibitem{Kiselev:1999sc}
  V.~V.~Kiselev, A.~K.~Likhoded and A.~I.~Onishchenko,
  Nucl. Phys.  B {\bf 569}, 473 (2000).

\bibitem{Kiselev:2000pp}
  V.~V.~Kiselev, A.~E.~Kovalsky and A.~K.~Likhoded,
  Nucl. Phys.  B {\bf 585}, 353 (2000).

\bibitem{Kiselev:1994ay}
  V.~V.~Kiselev,
  Mod. Phys. Lett.  A {\bf 10}, 1049 (1995).

\bibitem{Kiselev:1993eb}
  V.~V.~Kiselev,
  Int. J. Mod. Phys.  A {\bf 9}, 4987 (1994).

\bibitem{Kiselev:1992tx}
  V.~V.~Kiselev, A.~K.~Likhoded and A.~V.~Tkabladze,
  Phys. Atom. Nucl.  {\bf 56}, 643 (1993),
  [Yad. Fiz.  {\bf 56}, 128 (1993)].

\bibitem{Skwarnicki}  T.~Skwarnicki on behalf of LHCb Collaboration, http://cdsweb.cern.ch/record/1368193.

\bibitem{Kiselev:2002vz}
  V.~V.~Kiselev,
  arXiv:hep-ph/0211021.

\bibitem{Wang:2008xt}
  W.~Wang, Y.~L.~Shen and C.~D.~Lu,
  Phys. Rev.  D {\bf 79}, 054012 (2009).

\bibitem{Likhoded:2009ib}
  A.~K.~Likhoded and A.~V.~Luchinsky,
  Phys. Rev. D {\bf 81} (2010) 014015.

\bibitem{Likhoded:2010jr}
  A.~K.~Likhoded and A.~V.~Luchinsky,
  Phys. Rev.  D {\bf 82}, 014012 (2010).

\bibitem{EvtGen}
http://www.slac.stanford.edu/~lange/EvtGen.
 

\bibitem{Buchalla:1995vs}
  G.~Buchalla, A.~J.~Buras and M.~E.~Lautenbacher,
  Rev.\ Mod.\ Phys.\  {\bf 68}, 1125 (1996).

\bibitem{Anisimov:1998uk}
  A.~Y.~Anisimov, I.~M.~Narodetsky, C.~Semay and B.~Silvestre-Brac,
  Phys.\ Lett.\  B {\bf 452}, 129 (1999).

\bibitem{Huang:2008zg}
  T.~Huang, Z.~H.~Li, X.~G.~Wu and F.~Zuo,
  Int. J. Mod. Phys.  A {\bf 23}, 3237 (2008).

\bibitem{Schael:2005am}
  S.~Schael {\it et al.}  [ALEPH Collaboration],
  Phys.\ Rept.\  {\bf 421}, 191 (2005).

\bibitem{Kuhn:1990ad}
  J.~H.~Kuhn and A.~Santamaria,
  Z. Phys. C {\bf 48}, 445 (1990).
\end{thebibliography}
\end{document}